\documentclass{article}
\usepackage{graphicx}
\newcommand{\bea}{\begin{eqnarray}}
\newcommand{\beq}{\begin{equation}}
\newcommand{\eea}{\end{eqnarray}}
\newcommand{\eeq}{\end{equation}}

\begin{document}
\title{\bf Limits on the integration constant of the dark radiation
term in Brane Cosmology }
{\small
\author{A.S. AL-RAWAF\\
Physics Department, College of Science, King Saud University,\\
P.O. Box 2455, Riyadh, Saudi Arabia}
}
\maketitle
\begin{abstract}
We consider the constraints from primordial Helium abundances
on the constant of integration of the dark radiation term of
the brane-world generalized Friedmann equation derived from the
Randall-Sundrum Single brane model. We found that -- using simple,
approximate and semianalytical Method -- that the constant of
integration is limited to be between -8.9 and 2.2 which limits
the possible contribution from dark radiation term to be approximately
between -27\% to 7\% of the background photon energy density.
\end{abstract}
Although at present a fundamental quantum theory of gravity does
not exist, it is generally considered that string theory is the
leading candidate. But superstring requires ten extra-dimensions
or else bad quantum states become part of the spectrum. The Kaluza
-- Klein (KK) compactification was invoked to get rid of the superfluous
six extra dimensions where they were rolled up into some tiny
spaces of their own, until a new picture on extra-dimensions
emerged recently, such as in the Horova-Witten eleven-dimensional
super gravity. In this context, the ordinary matter fields are
not supposed to be defined every where but, in contrast are assumed
to be confined in a sub-manifold, called brane, a domain wall,
embedded, in a higher dimensional space (The Bulk). Braneworld
models are conceptually different from compactified KK models
because they don't attempt to derive nongravitational forces
from the gravitational oscillations in the extra dimensions,
on the contrary, if the extra dimensions are large, the gravitational
oscillations have to die out quickly in those directions, so
that we can't detect them. There are still KK modes of oscillations
in the extra dimensions, but because they couple through gravity,
and gravity is mostly confined to the bulk, they are effectively
invisible to our world on the brane. In this context an interesting,
string inspired proposal was given by Randall and Sundrum
\cite{ran}
where they considered only one extra dimension where our universe
is described as a three-brane embedded in a five-dimensional
anti-de-Sitter space AdS$_{s}$ \\
In this contest they proposed two models, one contains two branes
(RS1) which have equal and opposite tensions, and gives a new
approach to the hierarchy problem and a bulk that contains negative
cosmological constant. The other model propose only one brane
(RSII), with a positive tension, which may be thought of as arising
from sending the negative tension brane off to infinity (For
a Review see \cite{lang}). In this paper we discuss only the the single
brane model. The cosmological evolution of such a brane universe
has been extensively investigated and their solutions have been
found by several authors \cite{lang}-\cite{heb}. These solutions reduce to a generalized
Freidmann eq. on our brane which can be written as:
\beq
\left( \frac{\dot{R} }{R} \right) ^{2} =\left( \frac{K^{4} }{36}
\,\sigma ^{2} +\frac{\wedge _{5} }{6} \right) +\frac{K^{4} }{18} \sigma
\rho +\frac{K^{4} }{36} \rho ^{2} +\frac{E}{R^{4} }
\label{eq1}
\eeq
Where $\sigma$ is the brane tension and $\Lambda _{5} $
is the cosmological constant in the bulk, by fine-tuning the
brane tension and the bulk cosmological constant (which expresses
the brane cosmological version of the well-known cosmological
constant problem) the first term on the right hand side vanishes.
The second term can be identified with the density term in the
usual Friedmann eq. if we make the substitution
$8\pi G=\frac{K^{4} }{6} \sigma .$ For $\rho <<\sigma $,
this term dominates over the quadratic term
$\rho ^{2} $, which arises from the imposition of a function condition for
the scale factor on the surface of the brane, and it would decay
rapidly as $R^{-8} $ in the early radiation dominated universe. Hence it is expected
to be negligible in the low energy limit, and thus will not be
significant during the later nucleosynthesis and photon decoupling
epochs of interest here.

The last term which is known as the dark radiation term, and
it is derived from the electric part of the five-dimensional
Weyl tensor and carries information of the gravitational field
outside the brane \cite{shir}. The constant E is a constant of integration
and mathematically there is some justification for it to be either
positive or negative depending on the geometry in the bulk \cite{muk2}.
Our work here is to find the limits on the values of the constant
E from primordial nucleosynthesis of helium. The constraints
from Big Bang nucleosynthesis on the dark radiation term was
investigated in refs. \cite{ich} and \cite{brat}, where further in ref.
\cite{brat} they included the constraints on the 5-dimensional Planck mass
in the quadratic term. However our approach here is different
than the one presented in these two references, in that, we use
a semianalytical and approximate method with the observational
constraints to find the limits on the constant E, while in refes.
\cite{ich} and \cite{brat} they use solely the observational constraints.

Thus with the above considerations, eq.\ref{eq1} becomes (using c. g.
s. system):
\beq
H^{2} =\frac{8\pi G}{3} \,\rho _{r} +\frac{G\hbar }{c} \,\frac{E}{R^{4} }
\label{eq2}
\eeq
At temperature considered here, we can assume that $RT = \mbox{Constant}$.
We further normalize this constant to equal (in c.g.s system)
$\frac{hc}{K}.$ Thus eq.(\ref{eq2}) becomes:
\beq
H^{2} = \frac{8\pi G}{3} \,\rho _{r} +\left( \frac{G\hbar }{c} E\right) \,\left(
\frac{K}{\hbar c} \right) ^{4} \,T^{4}
\label{eq3}
\eeq
\bea
\mbox{Using}\;\; \rho _{r} =\left( \frac{ag}{2c^{2} } \right) \,T^{4}
\;\;\mbox{where}\;\;
a=\left( \frac{\pi ^{2} \,K^{4} }{15c^{3} \,\hbar ^{3} }
\right)\nonumber \\
\mbox{therefore} \,H^{2} =\frac{8\pi G}{3} \rho _{r} \,\left(
1+\frac{45}{4\pi ^{3} g} E\right) \nonumber\\
=\frac{8\pi G}{3} \rho _{r} \,\left( 1+0.03E\right) \,\,\,\,\,\,,
 \mbox{Using}
\;\;g=\frac{43}{4}
\label{eq4}
\eea

The primordial production of $^{4}$He is controlled by a competition
between the weak interaction rates and the expansion rate of
the universe. As long as the weak interaction rates are faster
than the expansion rate, the neutron-to-proton ratio
$\left( \frac{n}{p} \right) $
tracks its equilibrium value. Eventually as the universe expands
and cools, the expansion rate comes to dominate and
$\frac{n}{p} $
essentially freezes out at the so called freeze out temperature.
However the nucleosynthesis chain which begins with the formation
of deuterium through the process
$p+n\rightarrow D+Y$
is delayed past the point where the temperature has fallen below
the deuterium binding energy
$E_{B} $
since there are many photons in the exponential tail of the
photon energy distribution with energies
$E>E_{B} $
despite the fact that the average photon energy are less than
$E_{B}.$ (see for example \cite{olive}).

Thus the effects of the modification of density in eq.\ref{eq4} is to
increase or decrease (depending on the sign of E) the $^{4}$He abundance
through two factors, one by affecting the value of the freeze
out temperature, and the other by modifying the time available
for neutron decay. Use of the conservation eq. for radiation
with eq.\ref{eq4} leads to:
\beq
t_{m} =\left( 1+0.03E\right) ^{-\frac{1}{2} } \,\,t_{s}
\label{eq5}
\eeq
where $t_{m} $
is the modified time from freeze out to start of nucleosynthesis
and $t_{s} $
is the time calculated from standard model without the extra energy.

The calculation of primordial element abundances is a highly
nonlinear problem with many coupled nuclear reactions, and requires
a numerical analysis. There are many different numerical codes
for doing this calculation starting with Wagoner \cite{wag}. They mainly
differ in the different factors, they include in their calculations,
and particularly the different values they use for the neutron
half life \cite{yang,lop}. Here we present a simplified and approximate
method. In this approximation the primordial helium abundance
$Y $ is given by:
$\left(x =\frac{N_{n} }{N_{p} } \right):$
\bea
Y&=&\left( \frac{2x}{1+x} \right) _{F} \exp \left[ -\lambda \left(
t_{m} -t_{F} \right) \right]  \nonumber \\
&\cong &\left( \frac{2x}{1+x} \right) _{F} \exp \left[ -\lambda \,t_{m}
\right]
\label{eq6}
\eea
Where F represent freeze out and we ignored $t_{F} $
since it is of order one second. \\
Now using $x=\frac{N_{n} }{N_{p} } =\exp \left( -\frac{1.5}{T_{f} 10} \right) $
Where $T_{f10} $ is the freeze out temperature
$\left( T_{n} =\frac{T}{\left( 10^{n} K\right) } \right) $
, and is determined by equating the Hubble constant H to the
weak reaction rate $\eta $ for $n\leftrightarrow p$ conversions. Now
$$H\propto \,g^{\frac{1}{2} } \,T_{f10}^{2}
\,\,\,\,\,\,\mbox{and}\,\,\,\,\,\,\,\eta \propto T_{f10}^{5}$$
so that $T_{f}^{3} \propto g^{\frac{1}{2} } $ \\
Thus, if we assume that the change in density caused by the presence
of the dark radiation term is translated into a change in g.
Then if $g\rightarrow g+\delta g\,\,,\,\,\,x\rightarrow x+\delta x,$
$Y\rightarrow Y+\delta Y$, thus $\delta Y$ becomes
\bea
\delta Y &=&\frac{-x\,\,\ell n\,\,x\,\,e^{-\lambda \left( 1+0.03E\right)
^{-\frac{1}{2} } \,t_{s} } }{3\left( 1+x\right) ^{2} }
\,\,\,\,,\,\,\,\,\,\,\,\frac{\delta g}{g} \nonumber \\
\delta Y& = &\frac{-x\,\ell n\,x\,e^{-\lambda \left( 1+0.03\,E\right)
^{-\frac{1}{2} } \,\,t_{s} } }{\left( 3\right) \,\left( 1+x\right) ^{2} }
\,\,\,\,\,\cdot \,\,\,\,\,\left( 0.03\right) \,E
\label{eq7}
\eea
Using a predicted value calculated by Lopez and Turner \cite{lop},
where they found $Yp=0.246$,and taking a conservative value for the measured
value for $Y$ to be
$$0.23\,<\,Y\,\,<\,0.25$$
Which gives
$$-0.016\,\,\leq \,\,\delta Y\,\,\leq \,\,0.004$$
If E is positive, then the predicted value will increase, moving
closer to the observed high value, thus starting with the positive
value, putting $x=0.14$ (from$Y=0.246$),
$\lambda =78\times 10^{-5} $, and $t_{s} \cong 120\,\,s$
, eq.\ref{eq7} becomes
\bea
0.004 &=&\left( 0.002E\right) \exp \left[ -0.094\left( 1+0.03\right)
^{-\frac{1}{2} } \right]   \nonumber\\
&\cong& \left( 0.002E\right) \,\left[ -0.094\,\left( 1-0.015E\right)
\right]  \nonumber  \\
&\cong& 18.2\times 10^{-4} \,\left( 1+0.0014E\right)  \nonumber \\
\mbox{therefore} E &=&2.2 \nonumber
\label{eq8}
\eea
For the negative value, with
$\delta Y=-0.016$
, and noting the limit for negative E set by eq.\ref{eq5}, we obtain
$E=-8.9$.
therefore our main result is:
$-8.9\leq E\leq 2.2$
which limits the possible contribution from dark radiation term
to be approximately between $27\%$ and $7\%$ of the background photon
energy density. These limits are more restrictive than the values
found in ref.\cite{ich}, particularly for the negative limit. However,
we should further point out that the sign of the constant $E$ is
not yet a settled issue. In finding the exact cosmological, solutions
in the brane world, there were two approaches, one is to describe
the 3-brane as a 'domain wall' moving in the five dimensional
black-hole geometries \cite{kra,ida}. In this case and to avoid naked
singularity, the constant E must be negative, and its value is
limited by the relation $E\geq -\frac{K^{2} \ell ^{2} }{4} $
, where K is the curvature, and $\ell$ is the five dimensional curvature
length scale, which defines
the five dimensional cosmological constant through the relation
$\Lambda _{s} \equiv -\frac{4}{\ell ^{2} } $.
The other approach is by exactly solving the Einstein eqs.
In the Gaussian normal coordinate, where in this approach the
perturbation analysis constrain $E$ to be positive \cite{muk1}-\cite{heb}. Thus
there is no definite answer on the signature of this constant,
and some global non-linear analysis will be necessary to resolve
this issue.

\section*{Acknowledgement}
This research was supported by King Saud University, Research
Center Project No. (Phys/1424/01).

\end{document}